\documentclass{mn2e}
\usepackage{graphicx}
\usepackage[psamsfonts]{amssymb}

\def\eg{{\it e.g.\ }}
\def\ie{{\it i.e.\ }}
\def\etal{{\it et al.\ }}

\def\eg{{\it e.g.\ }}
\def\ie{{\it i.e.\ }}
\def\etal{{\it et al.\ }}

\def\msun{{\rm M_{\odot}}}

\begin{document}
\title[Brown dwarf populations in open clusters]
{Brown dwarf populations in open clusters}
\author[Adams, Davies, Jameson and Scally]{Tim Adams$^1$, Melvyn
B. Davies$^1$, Richard F. Jameson$^1$  and Aylwyn Scally$^2$  \\
$^1$ Department Physics \& Astronomy, University of Leicester, 
Leicester, LE1 7RH \\
$^2$ Institute of Astronomy, Madingley Road, Cambridge, CB3 0HA}

\date{Received ** *** 2000; in original form 2000 *** **}


\maketitle

\begin{abstract}
We present the results of multiple simulations of open clusters,
modelling the dynamics of a population of brown dwarf members. We
consider the effects of a large range of primordial binary
populations, including the possibilities of having brown dwarf
members contained within a binary system. 
We also examine the effects of various cluster diameters and masses.
Our examination of a
population of wide binary systems containing brown dwarfs, reveals
evidence for exchange reactions whereby the brown dwarf is ejected from
the system and replaced by a heavier main-sequence star. 
We find that there exists the possibility of hiding a large fraction
of the brown dwarfs contained within the primordial binary population.
We conclude that it is probable that the majority of brown dwarfs
are contained within primordial binary systems which then hides a large
proportion of them from detection.
\end{abstract}


\section{Introduction}

Brown dwarfs are essentially failed stars; they formed within 
a stellar nebula just like any other star, but they failed to reach a mass 
that generates sufficiently high central temperatures and pressures to induce 
the process of hydrogen fusion. As a result of the lack of fusion, brown 
dwarfs are naturally very dark objects. The little light that they do put 
out is normally at the Infrared end of the spectrum and is typically left 
over energy from the accretion process. 

	The identification of brown dwarfs is a difficult process;
their inherently faint magnitudes make them both difficult to locate
and to classify. Classification of brown dwarf status relies on the
identification of spectral features from the object which couldn't
have come from a low-mass star. The primary identifier of a brown dwarf
is the presence of a lithium resonance doublet at 6708 $\rm{\AA}$. This
feature up until recently proved very difficult to locate and the
existence of isolated brown dwarfs was a subject of some
controversy.

The first confirmed identification of a brown dwarf within the
Pleiades cluster was by Basri, Marcey \& Graham (1996) (Teide 1). They
successfully identified, spectroscopically, the lithium feature in a
brown dwarf candidate first discovered by Stauffer, Hamilton \& Probst
(1994). Since this first identification the assignment of brown dwarf
status has been given to many more candidate objects, particularly in
the Pleiades cluster.

	Due to its young age and proximity, the Pleiades cluster is an
excellent hunting ground for brown dwarfs. Many surveys have been made
of this cluster; Hambly \etal (1999) performed a survey in the I and R bands
and identified nine distinct single brown dwarf candidates. Further surveys 
have yielded more brown dwarf candidates within the cluster, notably a 
recent survey by Pinfield \etal (2000) has identified 30 possible brown 
dwarfs in a six square degree survey of the cluster. Of particular
interest is the work of Pinfield \etal (1998). 
Their examination of the dynamics of the Pleiades cluster lead them 
to believe that there are several thousand unseen brown dwarfs. 
If this is the case it is important to understand why so few brown
dwarfs have been found within the central portions of the cluster
which have been so well studied.

	The discovery of Gliese 229B (a brown dwarf contained within a
binary system) by Nakajima \etal (1995) (see also Golimowski \etal
1998, Basri \etal 1999) poses the
interesting possibility of containing cluster brown dwarfs within a
primordial binary population. Since this first discovery of a binary
containing a brown dwarf, a further six have been identified through
the use of data from the 2MASS survey (Gizis 2000, Skrutskie \etal 1995).
Mart\'{i}n \etal (2000) performed near-infrared photometry on very
low-mass members of the Pleiades cluster. They
failed to detect any resolved binary systems with a separation of more
than 0.2 arcsec; however they do manage to identify CFHT-P1-16 as a brown
dwarf binary of separation 0.08 arcsecs (equivalent 11 AU) by use
of HST data. They also detect
the presence of a binary second sequence within the colour magnitude
diagram (see Haffner \& Heckmann 1937, Hurley \& Tout 1998 for a discussion). 
However, they
conclude that there is a deficiency in the population of wide binary
systems (those with a separation greater than 27 AU); we consider this
issue in this paper.

Another cluster of interest is the Hyades. This is located at 
$\approx$46 pc from the sun, and is considerably older than the 
Pleiades ($\approx$650 Myrs as opposed to the Pleiades age of
$\approx$ 120 Myrs). Observations of this cluster reveal 
a deficit in low mass objects and brown dwarfs \cite{g99}, although
recent observations \cite{rm2000} have identified a binary system,
which may possibly contain a brown dwarf. However, 
in other regards it appears to be quite similar to the Pleiades, 
just older.

Work by Luhman \etal (2000) has demonstrated a strong similarity between the
Initial mass functions (IMF) of the Trapezium, Pleiades and M35 open
clusters. They performed sensitive, high-resolution imaging of the
central portion of the Trapezium cluster utilising the Near-Infrared
Camera and Multi-Object Spectrometer (NICMOS) aboard the Hubble Space
telescope, as well as performing ground based observations to take
K-band spectra for many of their sources. Their methodology allowed
observations of objects well below the hydrogen burning limit and so
we are now beginning to get a full explanation of the galactic field
IMF. 
Within the several hundred objects identified within the Trapezium
cluster, around 50 have been classified as brown dwarf candidates.
The derived Trapezium IMF is found to be similar to mass
functions predicted for other young star forming regions, \eg IC 348
(Luhman \etal 1998) and $\rho$ Oph (Luhman \& Reike 1999).
Indeed this work lends credence to the idea of a universal
IMF, at least in the case of open clusters (there appear to be
fundamental differences between these IMF's and those of globular
clusters). If this is true, a consistent model of brown dwarf dynamics
within a cluster should explain observed differences.

We seek to model the evolution of a cluster of stars which also
contains a population of brown dwarfs, in an effort to predict what
happens to the brown dwarf contingent that star clusters are predicted
to have. Stellar clusters may be modelled either through fokker-plank
codes or direct n-body integrators. It is this latter technique that
we apply. The use of n-body codes to simulate open cluster
evolution has become common place with more advanced codes allowing
more detailed study. The work of Terlevich (1987) demonstrated the use
of n-body simulations; she successfully modelled the
evolution of several clusters to their evaporation (\ie their total
dissipation) and examined the process of mass segregation within the cluster.

Earlier work by De la Fuente Marcos \& De la Fuente Marcos (1999) 
began the examination of brown dwarf evolution in open clusters. They
utilised the $\sc{Nbody 5}$ code by Sverre Aarseth and examined the
evolution of 8 separate cluster models which varied in their stellar
make up. They conclude for their models the relative percentages of
brown dwarfs to normal stars at older cluster ages is strongly
dependent on the IMF used at the start of their simulations. We seek
to further this work via the use of the more advanced code {\sc Nbody
6} also by Sverre Aarseth (see Hurley \etal 2001 for a review of the
{\sc Nbody 6} code). We examine the affects of various cluster
diameters, masses and density profiles. We also examine the implications
of various binary fractions and the effects they can have on a brown
dwarf population or at least appear to have.

The paper is divided into the following sections: In
section~\ref{sec:theo} we discuss the theoretical considerations
behind the simulations, detailing the important processes within the
cluster. In section~\ref{sec:initcon} we detail the various initial
conditions which were used for the simulations performed, within
section~\ref{sec:Numres} we outline the results of our various
simulations. These results are analysed in detail within
section~\ref{sec:dis}, before concluding remarks are made within
section~\ref{sec:concl}.

\section{Theory}
\label{sec:theo}

\subsection{Dynamics of the Cluster}
\label{sec:dyn}

The motion of the objects within the cluster leads to the definition
of two important time scales. The first of these time scales is the
cluster crossing time, t$_{\rm{cross}}$, which defines how long it
takes a star or brown dwarf to move across the cluster. It is defined
by the equation:

\begin{equation}
t_{\rm{cross}}=\frac{R_{\rm{hm}}}{\rm{\upsilon}}
\end{equation}

\noindent where R$_{\rm{hm}}$ is the half-mass radius of the cluster
and $\upsilon$ is the velocity dispersion.

Our second time scale is the relaxation time, t$_{\rm{relax}}$.
As the stars and brown dwarfs move within the cluster they will undergo 
gravitational interactions with each other.
The relaxation time refers to the period taken for a star to undergo 
sufficient interactions with various other bodies exchanging energy and have 
a resultant change in velocity of order ${\mid \delta
\underline{\upsilon} 
\mid = \mid \underline{\upsilon} \mid}$.  One may estimate the
relaxation time of a cluster based on the two-body relaxation time as 
defined in Binney \& Tremaine (1987):

\begin{equation}
t_{\rm{relax}}=\frac{N}{8 \ln N} t_{\rm{cross}}
\end{equation}

\noindent where N is the number of stars within the cluster.

	The exchange of energy during the two-body interaction is a very 
important driving force for the cluster. During an interaction between two 
stars, energy is transferred from the heavier to the lighter one,
until the cluster reaches a state of equipartition of energy. This results 
in the heavy star falling deeper into the cluster potential, namely toward 
the core of the cluster. This leads to the phenomena of mass segregation, 
whereby one finds the heaviest stars within a cluster migrating toward the 
core regions. Bonnell \& Davies (1998) demonstrate that the time scale for mass 
segregation 
within a cluster is well fitted by t$_{\rm{relax}}$ (as was predicted by Spitzer 1940). Hence systems which are 
older than their t$_{\rm{relax}}$ should be mass segregated.

	Whilst the heavy stars have lost energy during two body interactions, 
the lighter stars (or brown dwarfs) have gained energy. As a result the body's 
velocity naturally increases and so it can move further out into the cluster. 
After a sufficient number of interactions, it is possible that the
light star, or brown dwarf, 
may have a velocity which exceeds the escape speed of the cluster. This leads 
to the process of evaporation, whereby the cluster may lose mass via the 
escape of energetic stars.

\subsection{Binary population dynamics}
\label{sec:binpdyn}

Within the cluster environment there is likely to be a population of 
binaries. These will provide another important mechanism for 
driving the evolution of the cluster; interactions between binary 
systems and single stars provide another method of 
energy transfer within the system as we now briefly describe. 

There are two types of binary system, hard and soft. The definition of
hard and soft arise when a binary system undergoes an interaction with
a third star. We have to consider the ratio of the total kinetic
energy of the three bodies and the binding energy of the binary. If
the kinetic energy of the system is greater than the binding energy,
then there exists the possibility that energy can be passed into the
binary and cause it to break up; this is referred to as a soft binary
system. Whilst if the
kinetic energy of the three body event is less than the binding energy
of the binary, the binary is said to be hard. In this case energy
is transferred to the interloping star and the binary becomes tighter,
or harder. This transfer in energy then alters the energy budget of
the cluster. The dividing line between the hard and soft regimes occurs
when the total kinetic energy is just equal to the binding energy of
the binary and so leads to the definition of the critical velocity;

\begin{equation}
V_{\rm{crit}}^{2}=\frac{2GM_{1}M_{2}[M_1 +M_2 +M_3]}{M_3 (M_1 + M_2)}
\frac{1}{d}
\label{vcrit}
\end{equation} 

\noindent where M$_1$, M$_2$ are the masses of the 
two stars within the binary system and M$_3$ is the mass of the third
star, whilst d is the binary separation. If we equate the resulting
kinetic energy of the system to the binary system's binding energy, we 
can find the resulting definition of the hard soft boundary limit:

\begin{equation}
d_{h/s}=1744 \frac{M_1 M_2}{M_3 v^2 }\left(\frac{M_1 +M_2}{M_1 + M_2
+M_3}\right) \;\; \rm{ AU}
\label{dhs}
\end{equation}

\noindent where $v$ is now simply the relative velocity of the
interloping star in kms$^{-1}$ 
and the masses are  in solar units. Clearly we now see that the 
hard soft boundary of a star is now a function of the interloping 
star's mass. Thus a system might be hard to one interaction whilst 
being soft to another one.

As already mentioned, if a hard binary system were to undergo an interaction, 
it is expected to get harder; in doing so energy has to be transferred 
from the binary system to the third star. This results in an increase 
in the third star's velocity and can ultimately lead to its
evaporation from the cluster. Alternatively, if the interloping star
has a mass greater than one of the components within the binary, then
the two may be exchanged, with a hardening of the new binary system. A
trivial calculation, based in a binary system with components of 0.6
$\msun$ and 0.05 $\msun$, undergoing an interaction with a 0.4 $\msun$
star and forming a new binary which is $\approx$20 per cent harder
than the original system results in a kick velocity to the 0.05
$\msun$ body (a brown dwarf) of 4.9 kms$^{-1}$. The escape velocity of
our clusters is of order 2.5 kms$^{-1}$, so clearly if such an
interaction were to take place, the ejected brown dwarf would soon
escape from our cluster.

The interaction time scale for bodies within the cluster may be defined as:

\begin{equation}
\tau = \frac{1}{n\sigma v}
\label{rmaxpri}
\end{equation}

\noindent where $n$ is the number density of stars, $v$ is the
relative stellar velocity and $\sigma$ is the interaction cross-section.
The clusters within our simulations all initially have a constant
velocity dispersion, which is allowed to evolve with the cluster. 
As a consequence the interaction cross-section may be estimated as:

\begin{equation}
\sigma = \pi r_{\rm{col}}^{2}\left(1 + \frac{G(M_1 + M_2 + M_3)}
{v^2 r_{\rm{col}}}\right)
\label{eqn:intxs}
\end{equation}

\noindent where $v$ is the relative velocity of the binary and
the interloping stars and $r_{\rm{col}}$ is the distance of closest
approach for the system.

As stellar clusters evolve they are subject to tidal forces from the
galaxy within which they reside. These forces will lead to
perturbations on the orbits of the stars within the cluster. For
simplicity we model the motion of clusters moving on a circular orbit
about the centre of our galaxy at a radius equivalent to that of the
sun from the galactic centre (R$_G$=8.5kpc), this yields Oorts
constants of A = 14.5 $\pm$ 1.5 kms$^{-1}$kpc$^{-1}$ and B = -12 $\pm$ 3
kms$^{-1}$kpc$^{-1}$. With the addition of tidal forces to the
calculations, the equations of motion for the stars within the cluster
become \cite{gh97};

\begin{equation}
\ddot{x} =F_x + 2\omega_G \dot{y} +3 \omega_G ^2 x 
\end{equation}
\begin{eqnarray}
\nonumber
\ddot{y}=F_y -2\omega_G \dot{x}
\end{eqnarray}
\begin{eqnarray}
\nonumber
\ddot{z}=F_z - \omega_G ^2 z
\end{eqnarray}

\noindent where $\omega_G$ is the angular velocity about the centre of
the galaxy defined by:

\begin{equation}
\omega_G = \sqrt \frac{G M_G}{R_G ^3}
\end{equation}

\noindent with M$_G$ is the mass of the galaxy contained within a
distance R$_G$.

One of the most obvious effects of the tidal field is the existence
of a tidal radius for the cluster. This is the point at which the
gravitational forces due to the cluster and the galaxy balance; it is
defined by:

\begin{equation}
r_{\rm{tidal}} = \left(\frac{GM_{\rm{c}}}{4A(A-B)}\right)^\frac{1}{3}
\end{equation}

\noindent where M$_{\rm{c}}$ is the total mass of the open cluster. 
Once past this radius a star is no longer considered to be bound to
the cluster and moves off to become a part of the galactic disk.
Throughout all the simulations the effects of an external tidal field 
on the cluster are included with tidal radii calculated with the 
Oort constants as measured in the solar neighbourhood.

\section{Initial Conditions}
\label{sec:initcon}

\subsection{Stellar population}
\label{sec:stelpop}

Stellar masses for the population of stars within the simulations were 
produced by two methods. The first method was to simply utilise a
catalogue of stellar masses for the objects currently present with in
the Pleiades. To this catalogue we added a population of 100 low-mass
stars which were simply produced by doubling up the population of low
mass bodies in the catalogue. This allowed us to model the well studied
inner portions of the cluster. The addition of a further 100 bodies
was decided on via numerical experiments that we performed. 

The second method of producing stellar masses was to use the mass
function by Kroupa, Tout \& Gilmore 1993. With this function we produced a
distribution of stars with an upper limit on mass of 10 $\msun$ and a
lower mass of 0.08 $\msun$ (the hydrogen burning limit).
Evolution of the stellar population was accomplished via the use of
fitting formulas ~\cite{eg89}. 
In using the IMF by Kroupa, Tout \& Gilmore (1993) we extended our
investigations by looking at clusters of different masses, with
stellar numbers ranging between 1000 and 3000.

In addition to examining two different mass profiles we also
investigate the effects of two different initial density
distributions. The first is the Plummer distribution pattern which is
commonly used in
N-body simulations due to the fact that it's simple and fairly
realistic. The second set of models examined the evolution of a
uniform spherical distribution, which is preferred by some authors 
(e.g.de la Fuente Marcos 1999) as a model for open
clusters.

\subsection{Brown dwarf population}
\label{sec:bdpop}

Within our simulations we added to the cluster a population of 1500
brown dwarfs, each of which
have a mass of 0.05 $\msun$ and positions and velocities determined 
in the same manner as for the stellar population.
Investigations were made into using a brown dwarf IMF and populations
with a constant mass other than 0.05 $\msun$; however the choice of
brown dwarf mass had very little effect on the evolution of the
cluster or of the brown dwarf populations themselves.

\subsection{Binary population}
\label{sec:binpop}

Within the simulations performed between 0 and 500 primordial binary 
systems were added to the cluster. 
The binary systems were composed of stars all ready contained within
the cluster, thereby conserving the total mass of the cluster between
the simulations.
In the cases where the Pleiades masses have been used the components
of the binary systems have been randomly paired (Leinert \etal 1993,
Kroupa, Petr \&
McCaughrean 1999, Kroupa 2000 although for a differing view see for
example Mazeh \& Goldberg 1992). For the other
simulations, the IMF used produced the required binary components.
In each simulation a discrete fraction of the binary 
population was forced to have a brown dwarf as a secondary. 
Three numbers which result from this treatment are the fraction of
stars in binary systems, f$_{\rm{s}}$, the fraction of brown dwarfs contained
within binaries, f$_{\rm{bd}}$ and the fraction of objects (\ie brown
dwarfs and stars) contained within binaries, f$_{\rm{bin}}$:

\begin{equation}
\rm{f_{s}=\frac{N_b}{N_b + N_s}}
\end{equation}

\begin{equation}
\rm{f_{bd}=\frac{N_{bd,bin}}{N_{bd,bin} + N_{bd}}}
\end{equation}

\begin{equation}
\rm{f_{bin}=\frac{N_{b}+N_{bd,bin}}{N_{bd,bin} + N_{bd}+N_b + N_s}}
\end{equation}

\noindent where N$_{\rm{s}}$, N$_{\rm{b}}$, N$_{\rm{bd}}$ and
N$_{\rm{bd,bin}}$ are the 
number of single stars, the number of stars contained in binary 
systems, the number of brown dwarfs and the number of brown dwarfs 
contained in binary systems. The various fractions considered are listed
in Table ~\ref{binprops}. These were chosen so that we might investigate 
the effects that the different binary populations had on the evolution of 
the cluster. 

The positions of the binary systems were set to be consistent with the
distribution of stars in the particular cluster. The eccentricities of 
the systems were selected from a thermalised distribution (Jeans
1929), whilst the nodes and inclinations were randomly selected.

The separations of the binary components were chosen so that they 
were uniformly distributed in $\log$ d. This was accomplished using 
the following;

\begin{equation}
d=d_{0}10^{-x}
\end{equation}
\begin{equation}
x=A \log R
\end{equation}

\noindent where d$_0$ is the upper limit of the binary separation, A 
is a random number chosen from a uniform distribution between 0 and 1 
and R is a quantity known as the range. The range determines the 
spread in binary separations between the upper limit and d$_{0}$/R. A low 
value of R constrains the majority of the binary 
population to tight orbits whilst a high value leads to a greater spread in d.
Both scenarios of high and low R were examined for the clusters in
these simulations.

The separation of the binary components helps to determine what 
happens during a binary single encounter. When a tight binary 
undergoes an interaction, the energies involved tend to be much 
higher than during a corresponding interaction with a wide binary. 
However, the probability of a tight binary undergoing an interaction, 
is much lower than that of a wide system; this is simply  because 
it presents a much smaller cross section of interaction.  
To investigate the possible differences between the tight and 
wide systems, two distinct upper limits on the binary separation were
examined within the simulations. One, with d$_{\rm{0}}$=90 AU,
produced a tight population of binaries whilst the other,
d$_{\rm{0}}$=900 AU, produced a wider set (both with R=100).
Interactions involving the tighter binary population should lead to a
change in the energy makeup of the cluster. Either the population will
harden or the lighter member of the binary system (which could be a
brown dwarf) will be ejected with a substantial velocity which may be
sufficient for it to escape the cluster. Interactions involving the
wider binaries are more likely to result in the ionisation of the
binary system. The associated kicks given to the binary components
will be less than in the previous scenario, consequently it is possible
that brown dwarfs released from these soft systems remain within the
cluster. Our upper and lower limits on d allow us to investigate two
important scenarios and see what effect they have on the evolution of
the brown dwarf population.
Some argument could be made for selecting an even larger upper limit
for the separation of the binary components. Work by Gizis \etal (2001)
demonstrates that the population of very wide systems (d $>$ 1000 AU) is
non-negligible; however, our separations should allow us to investigate
the interesting effects within the cluster.

Table~\ref{binprops} details the properties of the binary
populations within the simulations we have run.

\begin{table*}
\begin{center}
\begin{tabular}{llllll} \hline\hline
\noalign{\medskip}
Model & d$_0$ & R & f$_{\rm{bin}}$ & f$_{\rm{s}}$ & f$_{\rm{bd}}$ \\
\noalign{\medskip}
\hline
\noalign{\medskip}
   I & 0.001 & 100 & 0.08 & 0.2, 0.18, 0.15, 0.13, 0.1 & 0, 0.017,
0.033, 0.05, 0.066 \\ 
  II & 0.001 & 100 & 0.16 & 0.4, 0.35, 0.3, 0.25, 0.2 & 0, 0.03, 0.06,
0.1, 0.13 \\
 III & 0.001 & 100 & 0.24 & 0.6, 0.53, 0.45, 0.38, 0.3 & 0, 0.05, 0.1,
0.15, 0.2 \\
 IV  & 0.001 & 100 & 0.32 & 0.8, 0.7, 0.6, 0.5, 0.4 & 0, 0.066,
0.133, 0.2, 0.266 \\
 V & 0.001 & 100 & 0.4 & 1.0, 0.88, 0.75, 0.63, 0.5 & 0, 0.083, 0.166,
0.25, 0.333 \\
   VI & 0.0001 & 100 & 0.08 & 0.2, 0.1 & 0, 0.066  \\	
 VII  & 0.0001 & 100 & 0.16 & 0.4, 0.2 & 0, 0.133   \\   
 VIII & 0.0001 & 100 & 0.24 & 0.6, 0.3 & 0, 0.2	  \\
 IX   & 0.0001 & 100 & 0.32 & 0.8, 0.4 & 0, 0.266  \\
 X    & 0.0001 & 100 & 0.40 & 1.0, 0.5 & 0, 0.333  \\
 XI   & 0.001  & 10  & 0.08 & 0.2, 0.1 & 0, 0.066  \\
 XII  & 0.001  & 10  & 0.40 & 1.0, 0.5 & 0, 0.333  \\
 XIII & -      & -   &  -   & -			  \\
\end{tabular}
\end{center}
\caption{Properties of the binary systems in some of the simulations.
 Here R is the range used within the simulation to determine the
binary properties, f$_{\rm{bin}}$, f$_{\rm{s}}$ and f$_{\rm{bd}}$ are the fractional
numbers of objects, stars and brown dwarfs contained within a binary
population.
The separations quoted are in model units.
}
\label{binprops}
\end{table*}

\subsection{Length scales in the N-Body code}
\label{sec:lengths}

Within the N-body code there exists a characteristic length scale
$\overline{\Re}$. This is a quantity which is fed into
the simulation at
the start and then all length scales during the simulation are scaled
by this value. This distance maybe linked to the characteristic length
scale associated with the Plummer model, b, which gives the space
density as a function of radial distance from the centre of the
cluster as:

\begin{equation}
\rm{ \rho(r) = \frac{3M_{c}}{4\pi b^{3}}\left(1
+\frac{r^2}{b^2}\right)^{\frac{-5}{2}}}
\end{equation}

via the formula ~\cite{a2001}:

\begin{equation}
\rm{\overline{\Re}} = \frac{16}{3 \pi}\rm{b}
\end{equation}  

During the simulations detailed values of $\overline{\Re}$ between 1.0 and 6.0
parsecs were investigated for both types of distribution pattern.

\section{Numerical Results}
\label{sec:Numres}

\subsection{Overall evolution of the cluster}
\label{sec:overall}

We begin by giving a brief overview of our results before giving a
greater discussion about each of the salient points.

Through our simulations we have found that the presence of a brown
dwarf population, regardless of their individual masses or numbers,
has a minimal impact on the evolution of the cluster as a whole. More
massive stars experience mass segregation toward the centre of the
cluster, whilst the lighter stars and brown dwarfs move to the outer
parts of the cluster.

As the lighter stars and brown dwarfs move outward, some fraction of
them gained a sufficient velocity to escape from the gravitational
potential of the cluster. Clusters which were initially more centrally
condensed (\ie those with small values of $\overline{\Re}$) had higher
velocity dispersions and as a result evaporated faster than less
tightly bound (\ie those clusters with large values of
$\overline{\Re}$).

For all the models investigated, we found that during the early part
of the cluster evolution (a few t$_{\rm{cross}}$) the escape rates of
brown dwarfs was virtually identical to that of the low-mass stars. At
later epochs some difference in the two rates would present itself;
however, this was dependent on the binary fraction of the simulation
as we shall discuss later.

We also found that both the initial density distributions evolve
toward a similar state.
That is
to say, that the clusters that we initially set up with a uniform density 
rapidly evolved (within a few t$_{\rm{cross}}$) to
a state similar to the equivalent Plummer model system. The major
difference between the two initial density distribution relates to the
number of bodies within the cluster at a given time. It was found that
the Plummer model tended to undergo an early phase of mass loss where
a fraction of the cluster was lost; however, this rapid loss only took
place for a short period of time. Past a few t$_{\rm{cross}}$ the loss
rates for the two distributions became essentially equivalent, which
is not overly surprising given the evolution of the uniform sphere and
the Plummer models toward a similar King Profile.

Examinations of various cluster sizes demonstrated that qualitatively
the evolution remains the same; however, the time scales over which the
processes occurred showed variation with the initial cluster properties.
Clusters with smaller values of $\overline{\Re}$ took a shorter period
(in years) to evaporate and tended to evaporate more uniformly, with
stars and brown dwarfs being lost at roughly the same rates.
For $\overline{\Re}$ values below four, the loss of brown dwarfs from the
Plummer and spherical model clusters were virtually identical;
however, for larger $\overline{\Re}$ values the Plummer models tended
to loose roughly twice as many brown dwarfs as the equivalent
spherical system at early cluster ages.

\begin{figure}
\begin{center}
\includegraphics[clip,width=0.95\linewidth]{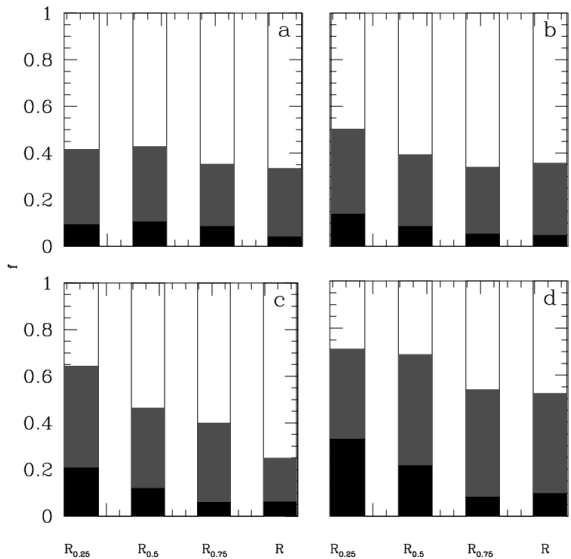}
\caption[]{Plots of how the populations of stars within a given region
change as a function of time. The above plots show how three different
mass bins contribute to the make up of a region. The black part of the
histogram represents the heavier stars M $\ge 1\msun$, the grey represents the
lighter stars of the cluster 0.05 $\msun <$ M $ <  \msun$ whilst the
plain area represents the brown dwarf population of the cluster. The
four regions considered were all fractions of the tidal
radius, R$_{T}$. Namely, 0 $\le$ R$_{0.25} < \rm{R_{T}/4}$,
$\rm{R_{T}/4 \le R_{0.5} < R_{T}/2}$, $\rm{R_{T}/2 \le R_{0.75} <3
R_{T}/4}$ and $\rm{3R_{T}/4 \le R < R_{T}}$. The four boxes represent,
(a) the initial dispersal, (b) the dispersal after $\approx$ 125 Myr
($\approx$ 10t$_{\rm{cross}}$, the age of the Pleiades), (c) $\approx$
300 Myr and (d) the dispersal after 650 Myr ($\approx$ 43
t$_{\rm{cross}}$, the age of the Hyades cluster).} 
\label{massseg}
\end{center}
\end{figure}

The first diagnostic commonly used to observe cluster evolution within
a simulation is to look at the process of mass segregation. Fig
~\ref{massseg} shows how three different mass bins contribute to the
make up of four shells within the cluster, as a function of
time. The three mass bins considered were 0-0.05 $\msun$,
0.05-1.0 $\msun$ and M $> 1 \; \msun$. As can be seen, initially all
three mass bins are evenly distributed within the cluster; however, mass
segregation is seen to take place rapidly.

	By the time the cluster has evolved to the age of 
10 t$_{\rm{cross}}$ (which in the case of a cluster with
$\overline{\Re}$=4.5 is 125 Myr, the age of the Pleiades cluster), we
already see an increase in the
fractional number of large (\ie M $\ge \msun$) mass stars within the
inner parts of the cluster, with a corresponding decrease in the
fractional numbers of brown dwarfs. Whilst in the outer parts of 
the cluster there is a build up in the number of brown dwarfs.

	This pattern continues on throughout the lifetime of the
cluster. The last of the four plots represents the cluster when it
has reached an age of $\approx$ 43 t$_{\rm{cross}}$. Here we note that the
fractional number of heavy stars appears to increase over the entire
cluster; what is really happening is that the lighter stars
and brown dwarfs are being lost preferentially to the heavy stars and
so the fraction of heavy stars within the cluster, as a whole, is seen
to increase.

\begin{figure}
\begin{center}
\includegraphics[clip,width=0.95\linewidth]{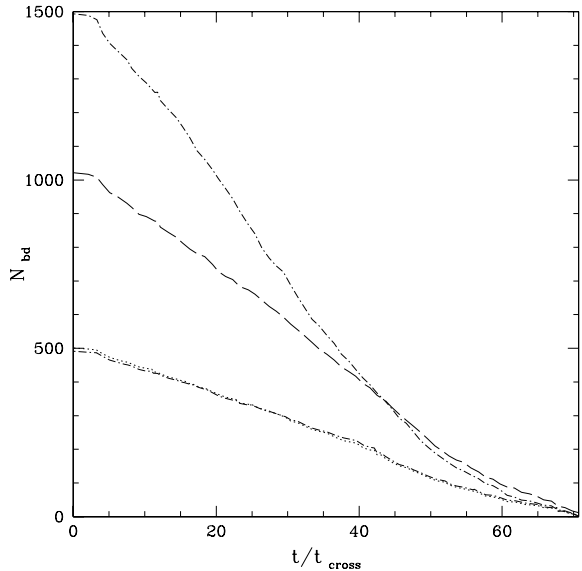}
\caption{Evolution of the populations of the stellar and brown dwarf
populations within a simulation that contained 500
primordial binarys each of which had a brown dwarf as the
secondary, \ie a simulation where f$_{bd}$ = 0.333 (see
section~\ref{sec:binpop} for a full definition of f$_{bd}$).
The dotted with long-dash line represents the single population of
brown dwarfs, the long dashed line represents the population of single
stars. The dotted line represents the brown dwarf population which is
contained within a binary system with another cluster member and
finally the dotted with short-dash line represents the stellar
population contained within binary systems. Note that the number of
objects in this figure are cumulative, \ie there are 500 brown
dwarfs tied up in binary systems and a further 1000 free brown dwarfs
at the start of the simulation.}
\label{BDSTARcomb}
\end{center}
\end{figure}

Fig~\ref{BDSTARcomb} shows how the number of
brown dwarfs and stars contained within the cluster varies as a
function of time for a simulation which initially contained 500 
primordial binary systems, each of which had a brown dwarf as a 
secondary. These two figures help us to understand the apparent
increase in heavy star population present in Fig~\ref{massseg}. At
the start of the simulation the normal stars are outnumbered 3:2 by
brown dwarfs, consequently the heavy star population initially makes
up a very small fraction of the entire population. However, when the
cluster is $\approx$ 35 t$_{\rm{cross}}$ old we see that there is a near
1:1 relation between the brown dwarf and star numbers. Thus the
fractional number of heavy stars within the cluster will have increased.

Another interesting feature that we note in Fig ~\ref{BDSTARcomb}, is
a ``$\it{repletion}$'' effect in the single body 
population of the cluster at around 40 t$_{\rm{cross}}$. At this 
point, a number of binary systems are broken up leading to a 
decrease in the binary population of the cluster, but an increase 
in the number of single bodies. 
Thus there exists a 
method of re-populating the brown dwarf population in a cluster. 
In the case detailed, the number of freed brown dwarfs is fairly 
low. We shall discuss this effect in more detail in section 5.2.

Another useful diagnostic of cluster evolution is to look at the variation 
of the core radius. This may be achieved via the equation of 
Casertano \& Hut (1985);

\begin{equation}
\rm{R_{c}^{2}=\frac{\Sigma^{N_{20}}_{i=1} R_{i}^{2}\rho_{i}^{2}}
{\Sigma_{i=1}^{N_{20}} \rho_{i}^{2}}}
\end{equation}

\noindent where R$_{\rm{c}}$ is the core radius, R$_{\rm{i}}$ is the
radius to the 
ith star in the summation and $\rho_{\rm{i}}$ is the mass density around 
the ith star not including itself. The summation is performed over the 
inner most 20 per cent of bodies. This relates, to within 6 percent,
to the observed core radius as defined by King 1962.

\begin{figure}
\begin{center}
\includegraphics[clip,width=0.95\linewidth]{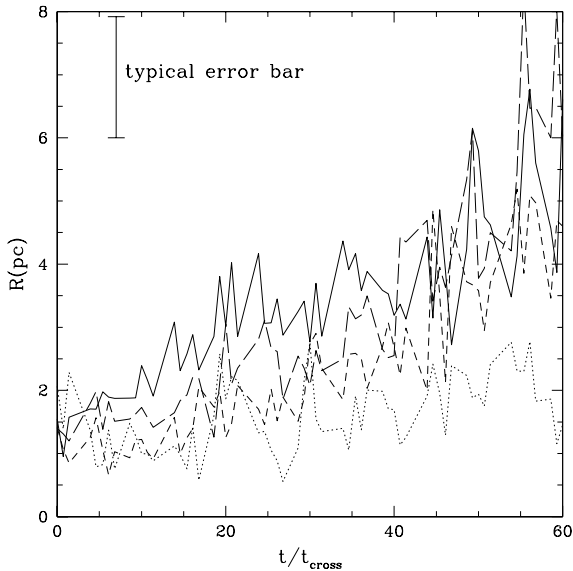}
\caption{The time averaged variation of the core radii (in pc) for the
brown dwarf population (solid line) and various stellar
populations. The dotted line 
represents the heaviest stars within the cluster, with a mass greater 
than 2 $\msun$. The shortest dashed line, represents stars with a mass 
between 1 and 2 $\msun$, whilst the longer dashed line represents stars 
with masses between 0.6 and 1 $\msun$. The rest of the stellar population
hasn't been plotted for diagram clarity. However, it followed the general 
trend of the low mass stars and brown dwarfs. Error bars in this and
all other plots are the standard deviation errors (1 sigma) between
different realisations of a particular set of cluster parameters. The
error bar plotted in this figure may be regarded as the typical error
on each data point. Variation from this value is less than 0.3 pc.}
\label{corerad}
\end{center}
\end{figure}

We performed a separate analysis for four distinct mass bins within
the cluster. These are shown in Fig~\ref{corerad}. The core radius of
the heavy stars (M $\geqslant$ 2 M$_{\odot}$ is denoted by the dotted line and
as can be seen it decreases
from its initial value as the stars sink toward the deepest parts of
the cluster potential. The brown dwarfs (solid line) in contrast have
a core radius which
increases with time. This corresponds to their outward motion,
following interactions with heavier bodies. Between these two extremes
we see that the overall trend is for an increased core radius, with
stars gradually moving out from the cluster centre before they
eventually evaporate from the environment.

\begin{figure}
\begin{center}
\includegraphics[clip,width=0.95\linewidth]{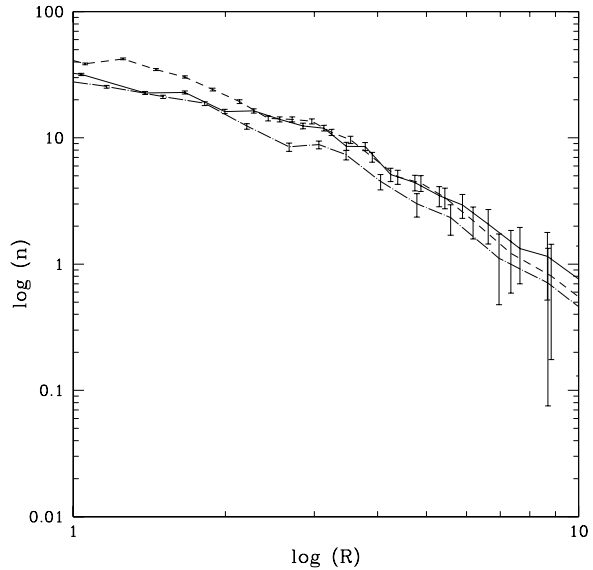}
\caption{A comparison of the brown dwarf surface density profiles
(number pc$^{-2}$) as a function of the cluster radius R (in pc) at
the age of the Pleiades. The
solid line represents the profile of a cluster which contained 500
primordial binaries, none of which had a brown dwarf member. The
dashed line represents the profile of a cluster which contained 500
primordial binaries, each of which had a brown dwarf member. The
dashed and dotted line is for the same cluster, but this time the
binary brown dwarf components have been removed from the analysis. }
\label{bdprofile}
\end{center}
\end{figure}

An important diagnostic for our investigation is to look at the surface density
 profiles of the brown dwarfs. Fig~\ref{bdprofile} shows the profiles for 
two simulations, once they had reached an age of 10 t$_{\rm{cross}}$
 (equal to the age of the 
 Pleiades). Both of these simulations contained 500 primordial binary systems;
 one set of binaries contained only stellar members, whilst the other 
simulation 
had a brown dwarf secondary in all of the systems. 
This second simulation is denoted by the dashed line on
Fig~\ref{bdprofile}. As can be seen, within the core regions of the
cluster the brown dwarf surface density is much increased in
comparison to the non-brown dwarf containing binary population.
However, when the brown dwarfs which were contained within the binary
systems are removed from the data and the profile is re-plotted, we see
 that the
density peak is much lower. This indicates that the brown dwarfs are
actually being dragged toward the core of the cluster by their 
heavier primaries. In essence the binary system may be regarded 
as a single body with a mass equal to the sum of its two components 
and an interaction cross section given by the binary separation. 
Thus we expect the systems as a whole to experience mass 
segregation and sink toward the deeper part of the cluster potential.

Another important quantity to consider within our simulations is the escape 
rate of both stars and brown dwarfs. It is found that the escape rates of 
stars remains largely unchanged by an increase in f$_{\rm{bin}}$ with 
typical values 
of between 10 and 35 per cent of the initial members lost by an age 
of 10 t$_{\rm{cross}}$. However, the loss rates of brown dwarfs do
show a  strong correlation to the values of f$_{\rm{bin}}$ and
f$_{\rm{bd}}$. If
we simply increase the value of f$_{\rm{bin}}$ keeping f$_{\rm{bd}}$
low, even
zero, then we see an enhanced ejection of brown dwarfs over that of
normal stars. However, if we increase f$_{\rm{bd}}$ at 
the same time, then this enhancement can be suppressed.

To understand the discontinuity in escape rates for the different
simulations, we must consider the time scale over which an interaction
between a binary system and a single body will lead to the exchange of
a body or the splitting up of the binary, t$_{\rm{enc}}$. For a binary
system that is at the hard-soft boundary, this time scale turns out to
be well matched by the local relaxation time scale, t$_{\rm{relax}}$; 
however,
for a binary system with a separation much less than d$_{\rm{h/s}}$
we find t$_{\rm{enc}}\gg$ t$_{\rm{relax}}$. Within our simulations, a
number of the brown dwarf containing binaries are hard to an
interaction with an intruder of 0.6 $\msun$. Hence, we expect the brown
dwarf binary population to exist within the cluster for a period
in excess of the relaxation time. This then explains why we see a
higher population of brown dwarfs within the high f$_{\rm{bd}}$ clusters at
later cluster ages.

We also find that the escape rates of both the stars and the brown
dwarfs are dependent on the initial cluster size and the cluster
distribution pattern. 
As is expected, clusters which had smaller values of $\overline{\Re}$
evaporated in a shorter period (in years) compared to clusters with larger
values of $\overline{\Re}$.
For clusters with an $\overline{\Re}$ less than four
the evaporation rates
of the Plummer and uniform  distribution patterns were almost
identical; however, for larger values of $\overline{\Re}$ it was found that the
Plummer models underwent an enhanced mass loss at early cluster
ages (10--15 t$_{\rm{cross}}$) relative to the same size uniform
distribution pattern. At later cluster ages the loss rates for the two
models become more even; however, the Plummer model is obviously more
depleted than the uniform model.

\subsection{Comparison to the Pleiades cluster}
\label{sec:plcomp}

\begin{figure}
\begin{center}
\includegraphics[clip,width=0.95\linewidth]{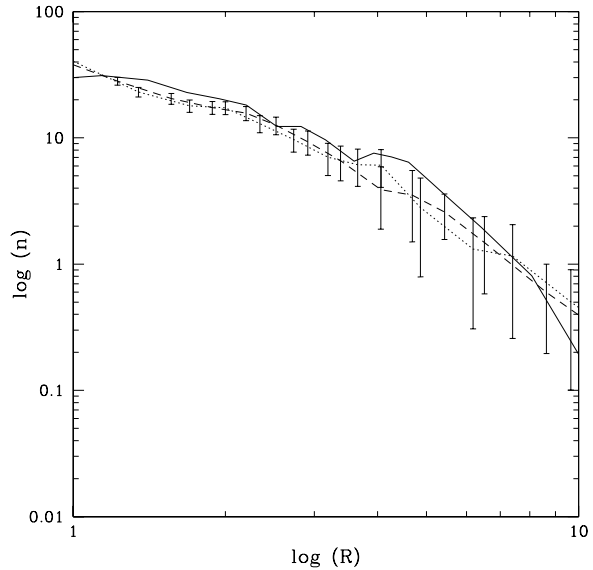}
\caption{A comparison of the surface stellar density (all masses), n,
as a function
of radius for the real Pleiades cluster (solid line) and two
simulations. Both the simulations contained 100 primordial binaries;
the dotted line had both members of the binary system as main sequence
stars, whilst the dashed line represents a cluster where all the
binary systems had a brown dwarf component. }
\label{pleiadcomp}
\end{center}
\end{figure}

A comparison can be made between the properties of some of our 
simulations and the  Pleiades cluster.
We make a comparison between the Pleiades and our simulations which
had an $\overline{\Re}$ of 4.5 parsecs.
 Perhaps the most obvious 
comparison is shown in Fig~\ref{pleiadcomp}.
In this figure we compare the stellar surface number density 
profile of the Pleiades cluster to that of a simulation 
containing 100 binary systems, each of which was made up of 
two stars, and another simulation which also had 100 binary 
systems but with each of these containing a brown dwarf as a 
secondary. As can be seen, over the majority of the cluster, 
the three profiles match remarkably well, with the only region 
where the simulations show an appreciable difference being the 
core.

In Table~\ref{pleiadprops} we make a comparison between the properties
of some of our simulations and data for the Pleiades cluster.
As can be seen, we closely match many of the physical parameters
describing the cluster, in particular the core radius and the crossing
times.

\begin{table*}
\medskip
\begin{center}
\begin{tabular}{p{2.0truecm}p{2.0truecm}p{2.0truecm}p{2.0truecm}} \hline\hline
Property & Real Pleiades cluster & Simulated cluster with no brown 
dwarfs & Simulated cluster with 1000 brown dwarfs \\
\hline
No. of stars & $\approx$ 900  & 1013  & 920 \\ 
Tidal Radius (pc) & 13.1  & 12.15  & 13.05 \\ 
Core Radius (pc) & $\approx$0.91  & 0.92  & 0.945 \\ 
Crossing Time (Myrs) & 15  & 15.58  & 14.756 \\ 
\hline
\medskip
\end{tabular}
\end{center}
\caption{Table of the physical quantities of the real Pleiades 
cluster, along with two of the computer simulations when they had 
reached a similar age. Note that the core radius for the Pleiades 
cluster is actually that for stars with masses between 3 and 
$12 M_{\odot}$ and so is slightly less than the figures quoted 
for the simulated environments, which look at the cluster as a 
whole.} 
\label{pleiadprops}
\end{table*}

\subsection{Future evolution}
\label{sec:futev}

Evolution of the cluster beyond the age of the Pleiades ($\approx$10 
t$_{\rm{cross}}$; for $\overline{\Re}$ = 4.5 this is equivalent to 125 Myr)
is shown on
many of the figures presented in this paper. For instance in
Fig~\ref{massseg} we show the continuation of mass segregation.
The lower right hand figure shows mass segregation in the cluster
at an age equivalent to the Hyades ($\approx$ 43 t$_{\rm{cross}}$
$\approx$ 650 Myr). 
As can be seen, by this epoch many
of the stars within the inner regions of the cluster have a mass
greater than 1 $\msun$, whilst the brown dwarf population has primarily been
exiled to the outer regions of the cluster.

Although we see an increased fractional number of heavy stars 
within the core regions of the cluster, we note that the total 
population of all bodies within this region actually declines 
over time. As time progresses, we are typically left with a few 
binary systems, containing relatively heavy stars in tight orbits, 
in the core region, whilst the rest of the stellar 
population moves outwards and eventually evaporates from the 
cluster.

Our most important result is regarding the evaporation of the brown dwarfs
from the cluster. For the first few crossing times (10---15) the evaporation of
brown dwarfs had been fairly similar to that of the lower mass stars
within the simulation, with a loss of both types being in the region
of 5 to 35 per cent. However, once past this point the
ejection rate of the brown dwarfs may drastically increase. By
50 t$_{\rm{cross}}$ we observe differences between the two escape rates of
up to 25 per cent in our most extreme simulations. This increased escape rate 
is linked to the interaction between the binary population 
and the single bodies in the cluster. Virtually all the binaries will 
be hard to an interaction with a brown dwarf and so the result of a 
brown-dwarf binary encounter will be an increased velocity for the 
brown dwarf. This increased velocity will eventually lead to the 
evaporation of the brown dwarf from the cluster. 
We never completely lose all the brown dwarfs from
the cluster; however, their numbers are
drastically reduced in comparison to the stellar content.

\begin{table*}
\medskip
\begin{center}
\begin{tabular} {lrcrrrrrcp{4truecm}} \hline \hline
Name & d & t$_{\rm{cluster}}$ & t$_{\rm{cross}}$ & t$_{\rm{rel}}$ & M
& r$_{\rm{tide}}$ &
r$_{\rm{core}}$ & f & references\\
\hline
NGC 2516 & 373 & 110 & 9 & 220  & 1000 & 13.0  & - & 0.89 & Abt \& Levey
(1972), 
Dachs \& Kabus (1989), Hawley \etal (1999) \\
Pleiades & 135 & 125 & 15 & 90  & $\approx$ 1000 & 13.1 & 0.91 & 0.88 & 
Pinfield \etal (1998) \\
NGC 2287 & 655 & 160 - 200 & - & - & $\ge$120 & 6.3 & - & 0.78 - 0.84 & 
Harris \etal (1993), Ianna \etal (1987), Cox (1954)\\
Praesepe & 174 & 400 - 900 & 19.4 & 370 & 1160 & 12.0 & 2.8 & 0.56 - 0.02 & 
Andrievsky (1989), Jones \& Shauffer (1991), Mermilliod \& Mayor
(1999), 
Mermilliod \etal (1990), Hodgkin \etal (1999)  \\
Hyades   & 46  & 625 	  & 18 & 390  & 500 - 1000 & 10.3 & 2.6 & 0.1 -
0.27 & Perryman \etal (1998), Reid \& Hawley (1999)  \\
NGC 2660 & 2884 & 900 - 1200 & 22.8 & 315 & $\ge$400 & 9.6 & 1.5 & 0.08 -0 & 
Frandsen \etal (1989), Hartwick \& Hesser (1971), Sandrelli \etal (1999) \\
NGC 3680 & 735 & 1450 & 7.5 & 28 & $\ge$100 & 4.3 & 0.6 & 0 & 
Hawley \etal (1999), Nordstr\"{o}m \etal (1997, 1996) \\ 
\hline
\medskip
\end{tabular}
\end{center}
\caption{Table listing the physical properties of some of the nearby
open clusters. The individual columns describe; the distance to the
cluster in parsecs, d; the age of the cluster in Myrs,
t$_{\rm{cluster}}$; the half mass relaxation time of the cluster in
Myrs, t$_{\rm{relax}}$; estimated total mass of the cluster in
$\msun$, M; the tidal radius in parsecs, r$_{\rm{tide}}$; the core
radius in parsecs, r$_{\rm{core}}$. The penultimate column lists the
fractional number of brown dwarfs left within our simulated clusters
when they had reached an age equivalent to that of the real
cluster. Whilst the final column gives references to papers where
information has been gathered from. Columns 2 through 7 adapted from
Portegies Zwart \etal (2000).}
\end{table*}

\section{Discussion}
\label{sec:dis}

\subsection{Evolution of the Cluster}
\label{sec:evclu}

We have observed that mass segregation takes place within the
cluster, with the brown dwarfs being moved out of the clusters central
regions, to be replaced by heavier stars. Along with this mass 
segregation, we have also observed an enhanced escape rate for the
brown dwarfs, compared to the stellar population, at later cluster ages. 

This preferential escape rate, at older cluster ages, may be very important 
in explaining the lack of brown dwarf observations in older clusters, 
such as the Hyades, which is much closer to the Earth than the Pleiades, 
although this is not the only explanation for a lack of observations of 
brown dwarfs in old clusters, as we shall discus later. The loss of many 
of the brown dwarfs relatively early in the 
evolution of the cluster, would allow them to move many 
degrees away from the cluster of interest. A trivial calculation shows 
that a velocity of 1 kms$^{-1}$, over a period of a million years, 
leads to a displacement of 1 pc. This being the case, if a brown dwarf
is to be within, say, four tidal radii (which is $\approx$ 13 pc for
the Pleiades cluster) then
it can't have left the cluster any more than 52 Myrs ago.

\subsection{Evolution of the binary population}
\label{sec:evbin}

Within the cluster, some of the most important interactions
will be between single stars and binary systems. 

When a single star encounters a hard binary system the binary 
becomes harder and the released potential energy is converted 
to kinetic energy, for both the binary system and the single 
star. The kinetic energy given to the single star may result 
in it having a velocity which exceeds the escape velocity of 
the cluster; if this happens then the single star will 
eventually evaporate from the cluster.

An examination of the simulation data revealed that relatively few 
binary systems had undergone any significant interaction by the age 
of the Pleiades. A few systems did demonstrate signs of softening 
and a few had undergone 
an exchange interaction, whereby the lighter star in the binary was 
exchanged for a heavier intruder. The number of interactions and 
exchanges were dependent on the upper size of the binary, 
d$_{o}$, and the range, R, used. The number of exchanges was also 
dependent on the brown dwarf binary fraction f$_{\rm{bd}}$.

To understand why only a few binaries have undergone interactions 
by the age of the Pleiades, one needs to examine equations 
(\ref{rmaxpri}) and (\ref{eqn:intxs}). By rearrangement of 
these two equations and substitution of the cluster age, $\tau$, 
one may obtain an expression for the maximum size of a binary 
that has yet to undergo an interaction;

\begin{equation}
R_{max}=\frac{1}{2}\left( - \frac{GM_{tot}}{v^2} +
\sqrt{\left( \frac{GM_{tot}}{v^2}\right)^2+\frac{4}{nv  \pi \tau}}\right)
\label{rmax}
\end{equation}

\noindent where, M$_{\rm{tot}}$ = M$_1$ + M$_2$ + M$_3$
and $v$ is the relative velocity between the binary and the 
intruder star. As with the hard-soft boundary, we see that this is a function 
of the intruder mass. This formula assumes we have a
constant velocity dispersion throughout the cluster. As the cluster 
evolves, this will no longer be the case and the formula will need to be 
modified. Any binary with a separation less than R$_{\rm{max}}$ will 
not have undergone an interaction and so its properties should remain 
unchanged.

\begin{figure}
\begin{center}
\includegraphics[clip,width=0.95\linewidth]{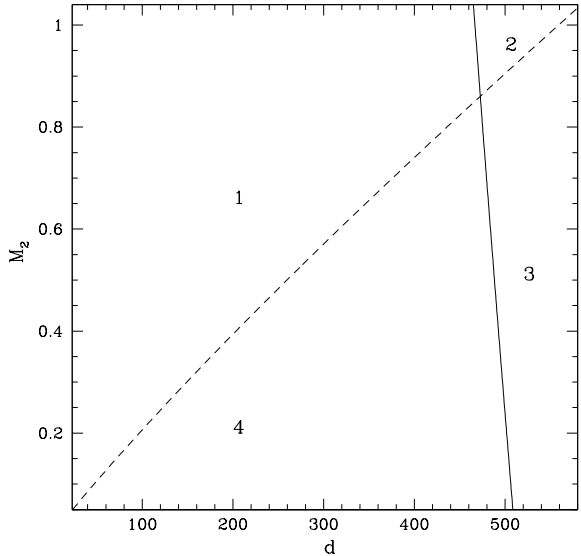}
\caption{Comparison of the hard-soft boundary (dashed line) of a binary
with the maximum size that a binary can have before it will interact
with another star (solid line) both in AU, for a variety of secondary 
masses (in $\msun$). The plot is for a primary mass of
1 $\msun$ with a velocity dispersion of 2 kms$^{-1}$. }
\label{dhssep}
\end{center}
\end{figure}

If we substitute into this formula for properties 
appropriate to the Pleiades cluster, n $\approx$ 150 pc$^{-3}$, 
M$_{1}$ = $\msun$, M$_{3}$=0.6 $\msun$ and $v$ = 2 kms$^{-1}$ we obtain 
Fig~\ref{dhssep}.

As marked, the figure may be broken into four distinct regions. The 
first region, region 1, extends from the y axis up to the solid line (which represents the maximum size of a binary yet to undergo an interaction) and lies
above the dashed line (which represents the hard soft boundary); binaries 
which occupy this region may be regarded as safe within the cluster 
environment. Statistically, they shouldn't undergo an interaction 
within the cluster environment. The second region, region 2, denotes hard 
binaries that will undergo interactions within the time allotted. 
These interactions will lead to either a hardening of the binary or 
an exchange of members. The third region, region 3, represents soft binaries 
which will undergo interactions. These binaries will either become 
softer or be broken up. The fourth region, region 4, represents soft binaries
which shouldn't interact within the period allotted.

\begin{figure*}
\begin{center}
\includegraphics[clip,width=0.95\linewidth]{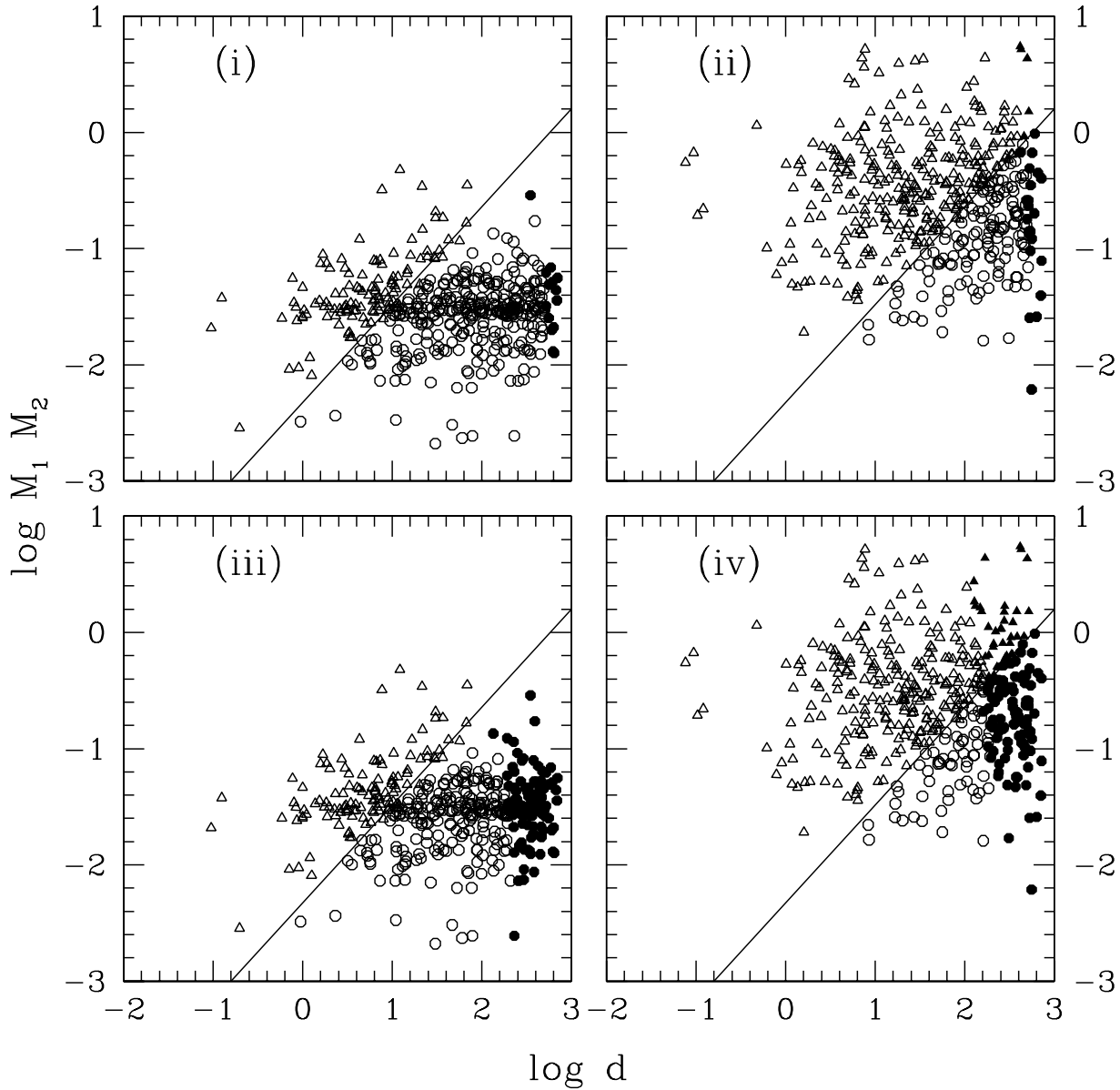}
\caption{Demonstration of how the binary systems within two
simulations may be regarded, in terms of
equations (\ref{dhs}) and (\ref{rmax}), at different times (d in AU,
M$_1$ and M$_2$ in solar units). The top row
represent clusters with an age similar to that of the Pleiades, whilst
the lower row represents clusters at the age of the Hyades. 
Two separate simulations were considered. Plots (i) and (iii) represent a
primordial population of 500 binary systems in which each of the
systems were forced to have a brown dwarf as a secondary; plots
(ii) and (iv) represent a population where the binary systems were only made
up of main sequence stars.
Open triangles represent binary systems which
may be regarded as hard and non-interacting within the allotted time
period, filled triangles denote binary systems which are hard and may
interact within the given period. Open circles represent
non-interacting soft systems and filled circles show soft interacting systems.}
\label{logplot}
\end{center}
\end{figure*}

We can investigate what equations (\ref{dhs}) and (\ref{rmax}) 
mean for the binary
systems within our simulations. Fig~\ref{logplot} shows a plot of the
binary separations against the product of the two components masses;
the different symbols used represent which region the binary systems
occupy on Fig~\ref{dhssep}. We show in this figure how the populations
of two different simulations may be regarded at two different epochs,
plot (i) and (ii) represent clusters at an age similar to that of the
Pleiades cluster, whilst plots (iii) and (iv) are results calculated for a
cluster with an age similar to that of the Hyades cluster. Plots (i) and
(iii) represent a simulation where a primordial population of 500 binaries
were forced to take a brown dwarf as a secondary, whilst plots (ii) and (iv)
are representative of a simulation where again 500 primordial binaries
were present but this time they were only made up of main-sequence stars.
Both populations of binary systems were set up with an upper limit to
their separations of d$_0$=900 AU and R= 100. 
Within these four plots it is possible to reconstruct the four regions
of Fig~\ref{dhssep}. Within Fig~\ref{logplot} the following symbols
have been used to denote what region a binary system would occupy in
Fig~\ref{dhssep}: open triangles represent region 1, filled
triangles denote region 2, filled circles are equivalent to region 3
and open circles denote region 4.
In each plot a line has been drawn; this may be
thought of as denoting the hard soft boundary line of
Fig~\ref{dhssep}. It is then possible to see that the four different
types of binary system are
clumped together in distinct regions. This indicates that the
interact-ability of a binary system is determined by its binding
energy, the higher the binding energy, the harder the binary, and the
less likely it is to interact. It can be seen that, the
number of binary systems that may be regarded as interact-able
increases with time, just as one would expect from
equation (\ref{rmax}). However, it is interesting to note that the majority
of systems may still be regarded as non-interact-able even after a
period of 650 Myrs.
\begin{table}
\medskip
\begin{center}
\begin{tabular} {ccccc} \hline\hline
symbol & plot i & plot ii & plot iii & plot iv \\
\hline 
$\circ$  & 340 & 155 & 246 & 86\\
$\bullet$ & 14 & 22 & 108 & 91\\ 
$\triangle$ & 146 & 318 & 146 & 285 \\ 
$\blacktriangle$    & 0 & 5 & 0 & 38\\ 
\hline
\end{tabular}
\end{center}
\caption{Numbers of the types of binaries present within the four
plots of Fig~\ref{logplot}. The symbols refer to the regions that the
binary systems would occupy on a plot similar to that of
Fig~\ref{dhssep}. An open triangle denotes region 1, a filled triangle
region 2, a filled circle region 3 and an open circle region 4.}
\end{table}

When a system was split up by an interaction with a third star, the
ejected star or brown dwarf typically had insufficient energy to 
escape from the cluster. As a result it in essence became a member 
of the free floating population subject to the same conditions as all 
the other free objects within the cluster.

The splitting up of a brown-dwarf containing binary, has interesting 
possibilities for re-populating the single brown dwarf contingent in 
the cluster. Within our simulations we have observed some evidence 
of ``$\it{repletion}$'' late in a clusters evolution; however the numbers 
involved are very small and not likely to make an appreciable 
difference in our ability to see free brown dwarfs within the cluster. 
In theory, we might be able to ``tune'' the binary separations, so 
that instead of following a description of being uniform in $\log$ 
d, they instead have a peak in separations which cause them to all
become inter-actable at roughly the same time. In doing so we 
might be able to induce a sizable ``$\it{repletion}$'' effect within the 
cluster. In actual observations this ``$\it{repletion}$'' effect would 
be seen as an increased brown dwarf population compared to what we 
predict should be present at a given time.

\subsection{Brown dwarf - brown dwarf binaries}
\label{sec:BD2}

Prompted by the discovery of several brown dwarf - brown dwarf binary
systems within the Pleiades cluster, a set of simulations were run
within which a distinct fraction of the brown dwarfs were forced to be
in a binary system with another brown dwarf, along with a population
of binary systems composed solely of stars. This fraction, which we
refer to as the double brown dwarf fraction, is defined so that;

\begin{equation}
\rm{f_{2BD}= \frac{N_{bd,bd}}{N_{bd,bin}+N_{bd}}}
\end{equation}

\noindent where, $N_{bd,bd}$ is the number of brown dwarfs contained
within a binary system with another brown dwarf. In all the
simulations we kept the total number of binary systems constant.

\begin{table}
\medskip
\begin{center}
\begin{tabular}{llll} \hline\hline
d$_0$ & R & f$_{\rm{bin}}$ & f$_{\rm{2BD}}$ \\
\hline
0.001 & 100 & 0.40 & 0.16 \\
0.001 & 100 & 0.40 & 0.33 \\
0.001 & 100 & 0.40 & 0.50 \\
0.001 & 100 & 0.40 & 0.66 \\
\hline
\end{tabular}
\end{center}
\caption{Details of the additional runs performed (along with those
detailed in Table~\ref{binprops}), to investigate the
effects of a population of binaries made up solely of brown
dwarfs. Within the table, d$_{0}$ is the maximum separation between
the binary components, R is the Range of the binary orbits \ie it
defines the spread in binary orbits, see section~\ref{sec:binpdyn} for
full details. f$_{\rm{bin}}$ is the fraction of objects within the
cluster that are contained in binaries, and f$_{\rm{2BD}}$ is the
fraction of brown dwarfs contained within binaries with other brown dwarfs.} 
\end{table}
 
The overall evolution of the cluster isn't effected by the population
of brown dwarf - brown dwarf binaries; however, the number of
exchanges within the cluster is sensitive to the value of
f$_{\rm{2BD}}$. As the fraction of binaries containing only brown dwarfs
rises, the number of interactions where one member of a binary system
is replaced by a heavier interloper falls. This may initially seem
contradictory, after all the binding energy of a double brown dwarf
system is much lower than that of a similar system made up with a main
sequence star; however, the reason for the decreased exchange rate
lies within equation (\ref{rmax}). Within this equation we see that the
total mass of the system is important in determining the maximum size
of a binary which has not yet interacted. This mass term arises because
of the effects of gravitational focusing, whereby the mass of the
binary system actually draws the interloping star toward it. In the
case of a brown dwarf - brown dwarf binary system, the total mass of
the binary is relatively low and so the effects of focusing are
negligible. Consequently, the maximum size of a binary, made up only of brown 
dwarfs, not to have undergone an interaction at some epoch is much greater
than that of a system which contains a main sequence star. 

As a result we expect the brown dwarf containing binary population to
exist within the cluster environment throughout much of the clusters
lifetime. The brown dwarf - brown dwarf binary, presents possibly the 
best opportunity for detecting brown dwarfs. In such a system, we don't 
have to worry about glare effects since both components have roughly 
the same luminosity. Another advantage is that the pair of brown dwarfs 
will move onto the binary second sequence with a brighter apparent 
magnitude than a single brown dwarf, aiding their detection.

\subsection{Binary system visibility}
\label{sec:binvis}

If we are to detect the two components within a binary system, an
important consideration is their separation.
With an $\it{HST}$ angular resolution of 0.1 arcsec, one can trivially 
calculate the minimum binary orbit that we should be able to resolve; 
this turns out to be 13.5 AU for the Pleiades cluster which is 135 pc 
from the Earth. This is of course for a binary system 
lying perpendicular to the line of sight (\ie an inclination of 0 degrees.)
 The natural inclination of 
the system to our line of sight means that many systems which 
have a separation of greater than 13.5 AU are still unresolvable.

One can show that the fraction of circular binary systems visible despite the
effects of inclinations is described by the formula:

\begin{equation}
f(x)=\sqrt{ 1 - x^2}
\label{fobsx}
\end{equation}

\noindent where x is the ratio of the true binary separation to the
minimum separation required for resolution.

If one were to consider our wider binary population (\ie a maximum
binary separation of 900 AU) one can show that when observations are
made of a cluster at the Pleiades distance with ground based
telescopes (which have their resolution limited to 1 arcsec by seeing
effects in the atmosphere) approximately 65 percent of the systems
would remain unresolved into two distinct sources. The situation would
become better with the use of the HST with approximately 14 percent of
the systems remaining unresolved; however, the cost in telescope
resources means that this is most unlikely to take place.

In addition to the problems associated with inclination effects, there
is another method of hiding a brown dwarf from direct observation
within a binary system. The low luminosity of a brown dwarf means that
it is possible to hide it via the glare of the primary. A reasonable
estimate for the difference in fluxes that would prevent resolution of
two distinct components is of order 10:1. This problem can be overcome
to some extent by making observations within infra-red bands where brown
dwarfs emit much of their energy, for instance in the K band, thus
ensuring the highest ratio of fluxes possible. With the 10:1
constraint on the relative fluxes, it is possible to calculate the
required angular separation between the two components of a binary
system if they are to be resolved optically. For a low mass main
sequence dwarf as the primary, the minimum required angular separation
is of order 0.7'' for a heavy (0.07 M$_{\odot}$ brown dwarf and
is of order 2.35'' for a lighter (0.01 M$_{\odot}$), hence cooler,
brown dwarf. Whilst for a primary like the sun the minimum angular
separation for a heavy brown dwarf and the primary is 1.35'' and for
the lighter brown dwarf it is 2.65''. These angular separations can
then be interpreted in terms of the minimum physical separation
between the binary pair. 
Clearly in the case of our softer binary population (upper separation
of 900 AU) located at the distance of the Pleiades, observations with
a ground based telescope would then only resolve between 12 and 27
percent of the entire binary population.

Even if the binary systems are not directly resolvable, it is still 
possible to discover the existence of a population of binaries. 
The main method of detecting an unresolved population is to look 
for the existence of a binary second sequence in the clusters 
colour magnitude diagram as we shall discuss in the next section.

\subsection{Brown dwarf cooling \& the binary second sequence}
\label{sec:bdcool}

Brown dwarfs have no renewable source of energy; that which they radiate 
comes from reserves built up during the original accretion from the nebula 
and subsequent contraction. As a result, brown dwarfs are continually 
cooling down; this means that they are become gradually dimmer.

This cooling is important in hiding brown dwarfs from sight. There are
a number of papers describing models of brown dwarfs, for example
Baraffe \etal (1998). They find that a 0.05 M${\odot}$ brown dwarf
will have a magnitude in the I band of 19.56 at the age and distance
of the Pleiades cluster, this is still within the limits of most
modern surveys. However, if one were to take into account the spectrum
of brown dwarf masses resulting from the tail of the IMF one finds
that the Pleiades observations
,with a limiting I band magnitude of 20 (which corresponds to
a lower detectable brown dwarf mass of $\approx$ 0.04 M$_{\sun}$), will
fail to detect some  
50 per cent of the brown dwarf population (assuming brown dwarfs have
masses between 0.02 and 0.075 M$_{\sun}$ described by the IMF of
Kroupa \etal 1995). In the case of the older Hyades cluster ($\approx$ 650
Myr) the fraction of brown dwarfs missed by a similar survey would be
about the same, the greater brown dwarf age off set by the clusters
proximity to us. More recent surveys extend the depths of observations
through to I=22, in the case of the Pleiades and Hyades this should
result in only $\approx$ 35 per cent of the brown dwarf population
being missed.

The previous arguments can only be applied to isolated brown dwarfs. 
Those within a binary system will be subject to different conditions. 
As was mentioned in the previous section, the presence of a binary 
population, which isn't directly resolvable, may be betrayed by the 
existence of a second sequence on the colour magnitude diagram. 

A binary system has a luminosity equal 
to the sum of the two components, but a colour which is redder 
than the equivalent star of the same luminosity (Haffner \& Heckmann
1937, Hurley \& Tout 1998). This leads to 
the formation of a second sequence lying $\approx$ 0.75 mag above the
main sequence on the colour magnitude diagram. 
This relative height of the second sequence is true for binary systems
with non-extreme mass ratios ($q=M_2 / M_1$); however, in the cases
where brown dwarfs are the secondary, an extreme mass ratio is likely to
occur. These systems tend to move away from
the second sequence and occupy the gap between it and the main
sequence line. 

If the mass ratio is particularly extreme, then the
redness of the low mass companion is insufficient to move the system
onto the second sequence. In this case the system lies either
somewhere between the two lines or very close to the main
sequence. The most extreme mass ratio that one expects to find lying
on the second sequence is of order, q=0.33 (this is within the I, I-K
plane [Steele \& Jameson 1995]). If q is less than this, the
binary moves away from the second sequence. In the case where a brown
dwarf is the companion, a low value of q is quite probable, unless we
also have a fairly low mass primary.

Clearly even for a massive brown dwarf ($\approx 0.07$ M$_{\odot}$)
if the binary system is to lie on the second sequence then the primary
can be no more massive than $\approx$ 0.2 M$_{\odot}$
Within our simulated data we
find that from all the binary systems still present at the age of the
Pleiades between 2 and  6 per
cent fulfilled the condition of q $\ge$ 0.33; clearly the vast
majority of binary systems will not be detectable in this
manner. 
As we have previously
discussed, brown dwarfs cool as they get older. This
results in their magnitudes becoming fainter and as a consequence of
this, a system which was initially on the binary second sequence will
move away from it. Thus, in old clusters, it is possible that a
population of brown dwarfs contained in binaries may be entirely
hidden from view.

Observations of the Pleiades cluster~\cite{SJ95} have shown some
evidence of a second sequence at very low
stellar masses, thus indicating the existence of a brown
dwarf population in binary systems.

\subsection{Probabilities of finding a brown dwarf}
\label{sec:probfinding}

We have detailed many methods by which brown dwarfs are lost from
observations; these range from dynamical depletion to hiding the brown
dwarf in a binary system. 
What do these losses imply for the probability of observing a brown
dwarf within a cluster?

	It is trivial to find out how many brown dwarfs are left in
the various simulations at various epochs, a general trend shows that
the more brown dwarfs which were initially contained within a binary
systems the greater the number still contained within the cluster at
later epochs.
However, it is not merely a question
of how many brown dwarfs are present within the cluster, but rather
how many can be seen. As we have said, at later epochs there tends to
be a higher number of brown dwarfs in the clusters which have a high
f$_{\rm{bd}}$, this is because the brown dwarfs in binaries are
retained by the cluster. Unfortunately, as we have discussed, a high
proportion of binary systems are hidden via inclination effects. 
Despite this, one still expects to
be able to see more brown dwarfs in these high $f_{\rm{bd}}$ clusters at
very old cluster ages compared to clusters which had fewer brown
dwarfs initially contained in binary systems. 

\begin{figure}
\begin{center}
\includegraphics[clip,width=0.95\linewidth]{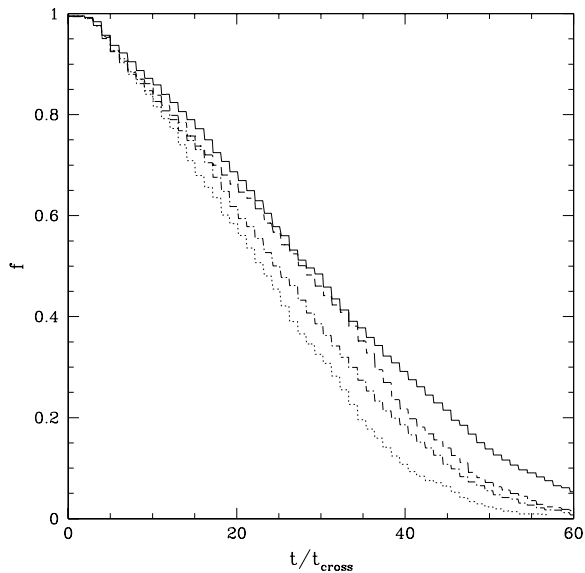}
\caption{The fraction of the original brown dwarf population still contained
with in the cluster as a function of the cluster crossing time for a
number of our simulations.
The solid line
represents a simulation containing 500 primordial binaries each of
which had a brown dwarf secondary, whilst the short dashed line is for
a similar simulation, except that this time no brown dwarfs were in
binary systems. The long dashed line represents a simulation
containing 100 primordial binaries, each of which has a brown dwarf
secondary, whilst the dotted and dashed line represents the same
simulation, except no brown dwarfs were contained within a binary
systems. All these simulations were performed with an
$\overline{\Re}$=4.5.}
\label{frac1}
\end{center}
\end{figure}

We see, in Fig~\ref{frac1}, that at
the age of the Pleiades ($\approx$ 10 t$_{\rm{cross}}$), the simulated
cluster still contains nearly
90 per cent of its original brown dwarf members. The most up to date
observations, however, have only revealed 30 brown dwarfs within the
real cluster (this is in part due to the fact that only a modest
fraction of the whole cluster has been surveyed to the required depth
to find brown dwarfs). When we look at a time closer to the Hyades age, 650
Myrs ($\approx$ 43 t$_{\rm{cross}}$), we see that within the more
pessimistic simulation where no brown dwarfs were contained within
binary systems only $\approx$ 10 per cent of the original brown dwarfs are
retained within the cluster.
The most
optimistic simulation indicated that, at the age of the Hyades, the
cluster would contain only $\approx$30 per cent of the original brown dwarf
population.

Combination of the data within Fig~\ref{frac1} and the various effects
which hide brown dwarfs (\ie cooling, being contained within a binary
system),  shows that
at the age of the Pleiades ($\approx$ 10 t$_{\rm{cross}}$) anywhere
between 40 and 50 per cent of the original brown dwarf  population
should, in principle, be viewable (this is with an I band magnitude
limit of 20, surveys limited to I=22 should be able to find up to 65
per cent of the brown dwarf population); whilst at the age of the Hyades
($\approx$ 43 t$_{\rm{cross}}$) this figure falls off to between 3 and
16 per cent. Thus, the probability of finding a brown dwarf in an old
cluster like the Hyades is very small.

\begin{table*}
\medskip
\begin{center}
\begin{tabular}{p{2.0truecm}p{2.0truecm}p{2.0truecm}p{1.5truecm}p{1.5truecm}p{2.0truecm}p{2.0truecm}} 
\hline 
   & Number of brown dwarfs originally & brown
 dwarfs still in the cluster & Number  observable in whole cluster& 
Number expected to be found in Pinfield \etal type survey &actual
 observation & Number of ``$\it{repleted}$'' brown dwarfs \\
\hline \hline 
binary systems & 500& $\approx$ 300   & 18 (54)  & 5(15) &10   & 20\\
single  & 1000 &$\approx$ 850 & 640  & 120 & 20 & \\
binary systems & 250&$\approx$ 200   & 12 (36) & 3(8) & 10   & 13\\
single & 1250& $\approx$ 1000  & 750  & 180 &20 & \\

\hline
\end{tabular}
\end{center}
\caption{A comparison of the observable number of brown dwarfs
contained within our simulated clusters and the current observations
of the Pleiades cluster. In the case of the binary systems we have two
figures quoted for the number of systems visible. The first
refers to those which should be found via the binary second sequence
as those found within the Pleiades have been. The second figure
refers to those which should be resolvable. In both simulations there
were initially 1500 brown dwarfs and 500 binary systems with a
discrete fraction containing
brown dwarfs (as detailed in the first column), the systems had an
upper separation of 900 AU. 
In the fifth column we detail the number of brown dwarfs that we would
expect to find within a survey
akin to that carried out by Pinfield \etal 2000. This survey covered a
total of six square degrees spread over five separate regions within
the central parts of the cluster.
As can be seen such a survey dramatically reduces the number of brown
dwarfs that we expect to see compared to a survey of the cluster as a
whole. We see that such a survey performed on our simulated clusters
would yield too few brown dwarf containing binary systems (found via
the second sequence), whilst it would find too many single brown
dwarfs when compared to observations of the real Pleiades cluster.}
\end{table*}

\section{Conclusions}
\label{sec:concl}

	Through our simulations we have been able closely to match the
observed stellar surface density distribution, as well as other key
properties of the Pleiades cluster, for a varied number of
brown dwarfs within the cluster environment. This indicates that the
presence of a brown dwarf population is unlikely to be betrayed by the
observed stellar population.
 
We
have demonstrated that, in the case of the Pleiades and Hyades clusters, 
the cooling of the brown dwarfs on their own
should be insufficient to hide them from view. 

However, if they are
contained within a primordial binary population, then the difference in
luminosity between the primary and brown dwarf companion might be
enough to make the companion unobservable, due to equipment
limitations. Observations lend credence to the existence of a 
primordial binary population within the Pleiades. In
particular the discovery of a binary second sequence by Steel \&
Jameson (1994) indicates that there are many unresolved binary
systems within the Pleiades cluster, all of which have the potential
to hide brown dwarfs. The work by Richichi \etal (1994) indicates that
around half of the observed stars in the sky must actually be
unresolved binary systems. This again lends credence to some of the
high binary fractions used in some of our simulations.

	Our multiple realisations of the cluster
have shown that the effects of different
brown dwarf binary fractions are minimal. The dynamics of the cluster
remain largely unchanged with key features such as the stellar surface
density profile and the loss rates of stars
remaining almost the same for a particular cluster size. 
Containing a contingent of brown dwarfs within a primordial binary
population has two key effects; first, the larger this contingent the
greater the population of brown dwarfs present within the cluster at
later cluster epochs. Secondly, by containing the brown dwarfs within
a binary system we have an effective method of hiding them from view,
both in terms of inclination effects and their continual cooling.

In the case of the Pleiades cluster we have demonstrated via the
combination of Figs~\ref{bdprofile} \&~\ref{frac1}, that
the effects of dynamical depletion of brown dwarfs is insufficient to
explain the low number of single brown dwarfs  observed in the well
studied central portions 
of the cluster ($\approx$ 30 strong  brown dwarf candidates  in
total, a third of which are in  binary systems with low mass companions).
However, in the case of the much older Hyades
cluster, depletion effects become far more important, with perhaps two
thirds, or more, of the initial brown dwarf population lost from the
cluster; and a greater proportion hidden from view by effects
associated with binary systems. This quite clearly helps to explain
why there are no confirmed brown dwarf sightings within the Hyades
cluster. 

Of concern, however, is the relatively low number of brown dwarf
observations within the well studied central portions of the Pleiades
cluster. The data from our
simulations would seem to indicate that the surface number density of
brown dwarfs within the central parts of the cluster would lend themselves
to detection.
This being the case, there
are a number of  possibilities for explaining the disparity between
observation and our results:

\noindent 1) The number of brown dwarfs within the cluster has been
over estimated. This seems likely as the work of Rabound \& Mermilliod
(1998) demonstrates the extreme uncertainties as regards the mass of
the Pleiades cluster. They use three distinct methods of predicting
the cluster mass; namely the use of the tidal radius, the virial
theorem and a proposed IMF, and get three different results with very
large confidence bands. 

\smallskip

\noindent 2) The cluster contains a population of very low mass brown
dwarfs. As a result of their low mass these brown dwarfs have quickly
cooled to a point below our detection threshold.

\smallskip

\noindent 3) A greater proportion of the brown dwarfs are contained
within moderately tight binary systems and so are not optically
resolvable. Brown 
dwarfs contained in systems with a low-mass primary (such that the
mass ratio of the system is greater than a third) will be detectable
via the binary second sequence on the CMD. Brown dwarfs contained in
systems with heavy primaries (M $>$ 0.2 M$_{\odot}$) may be detectable
via the radial shifts in the light from the primary, in much the same
way as the search for extra-solar planets is being conducted.

\smallskip

It is this later theory that we favour. The observed number of single
brown dwarfs compared to our predictions is so small it seems
unlikely that such a large single population exists
within the cluster. The effects of mass segregation cannot move a
sufficient fraction of the single brown dwarfs out of the well studied
central region of the Pleiades cluster to account for the low numbers
observed.
We have shown that the majority of brown
contained within a binary system are virtually undectable without a
massive search looking for radial shifts in the light of the
primary. 
We also see that our simulated data predicts too few binary systems
would be observed on the binary second sequence, this would strongly
suggest a much higher binary fraction within the cluster (which is of
course necessary to explain the main sequence star-star binary systems
we observe as well) providing a method of hiding a substantial
population of brown dwarfs within the cluster.
Further, an examination of our simulations demonstrates that a
considerable fraction of the currently single brown dwarfs may
actually have initially been in binary systems which were broken up,
\ie the single population is, in part, made up by the {\it repletion}
effects that we discussed in section~\ref{sec:evclu}. This would
remove the need for a large single population of brown dwarfs to be
present within the newly formed cluster.

\section*{ACKNOWLEDGEMENTS}
TA gratefully acknowledges support through a PPARC research student ship. MBD
gratefully acknowledges the support of the Royal Society through a
URF. We are grateful to Sverre Aarseth for providing us with a copy of
his N-body6 code.


\begin{thebibliography}{}
\bibitem[Abt \& Levy 1972]{al72} Abt H.A., Levy S.G. 1972, ApJ, 172,
355
\bibitem[Anderson 2001]{a2001} Anderson M. 2001 Masters Thesis,
Copenhagen University
\bibitem[Andrievsky 1998]{a98} Andrievsky S.M. 1998, A\&A, 334, 139
\bibitem[Baraffe \etal 1998]{b98} Baraffe I., Chabrier G., Allard F. \&
Hauschildt P. A\&A, 337, 403
\bibitem[Basri, Marcy \& Graham 1996]{bmg1996} Basri G., Marcy G.W. \&
Graham J.R. 1996, ApJ, 458, 600
\bibitem[Binney \& Tremaine 1987]{bt87} Binney J., Tremaine S., 1987, 
Galactic Dynamics, Princeton University press
\bibitem[Bonnell \& Davies 1998]{bd98} Bonnell I.A., Davies M.B.,
MNRAS, 295, 691 
\bibitem[Casertano \& Hut 1985]{ch85} Casertano S. \& Hut P. 1985,
ApJ, 298, 80 
\bibitem[Cox 1954]{c54} Cox A.N. 1954, ApJ, 119, 188,
\bibitem[Dachs \& Kabus 1989]{dk89} Dachs J., Kabus H. 1989, A\&As,
78, 25
\bibitem[de la Fuente Marocs \etal]{dfm99} de la Fuente Marcos 
R., de la Fuente Marcos C., 1999, Astrophysics and Space Science, 271, 127
\bibitem[Eggleton, Fitchett, \& Tout 1989]{eg89}  Eggleton P.P.,
Fitchett M.J. \& Tout C.A., 1989, Ap. J. 347, 998 
\bibitem[Frandsen, Dreyer \& Kjeldsen 1989]{fdk89} Frandsen S., Dreyer
P., \& Kjeldsen H., 1989 A\&A, 215, 287
\bibitem[Giersz \& Heggie 1997]{gh97} Giersz M. \& Heggie D.C. 1997
MNRAS 286 709
\bibitem[Gizis \etal 2001]{g2001} Gizis J.E., Kirkpatrick D.,
Burgasser A., Reid I.N., Monet D.G., Liebert J., \& Wilson J.C. 2001
Ap.J. 551, L163
\bibitem[Gizis 2000] {g2000} Gizis J.E. 2000, American Astronomical Society
Meeting 197, \# 127.01
\bibitem[Gizis \etal 1999]{g99} Gizis J.E., Reid I.N., Monet D.G.,
1999, AJ, 118, 997
\bibitem[Golimowski \etal 1998] {go98} Golimowski D.A., Burrows
C.J., Kulkarni S.R., Oppenheimer B.R., Brukardt R.A. 1998, AJ, 115, 2579
\bibitem[Haffner \& Heckmann 1937]{hh37} Haffner H., Heckmann O.,
1937, G\"{o}tt, Ver\"{o}ff., 55, 77
\bibitem[Hambly \etal 1999]{hh99} Hambly N.C., Hodgkin S.T., Cossburn M.R., 
Jameson, R.F., 1999, MNRAS, 303, 835 
\bibitem[Harris \etal 1993] {h93} Harris G.L., Fitzgerald M.P.V., Mehta
S., Reed B.C. 1993, AJ, 106, 1533
\bibitem[Hartwick \& Hesser 1971]{hh71} Hartwick F.D.A., Hesser
J.E. 1971, PASP, 83, 53
\bibitem[Hawley \etal 1999]{h72} Hawley S.L., Tourtellot J.G.,
Reid. I.N. 1999, AJ, 117, 1341
\bibitem[Hodgkin \etal 1999] {hod99} Hodgkin S.T., Pinfield D.J.,
Jameson R.F., Steele I.A., Cossburn M.R., Hambley N.C. 1999, MNRAS, 310, 87
\bibitem[Hurley \etal 2001]{htap2001} Hurley J.R., Tout C.A., Aarseth
J.A. \& Pols O.R., 2001, MNRAS, 323, 630
\bibitem[Hurley \& Tout 1998] {ht98} Hurley J., Tout C.A., 1998, MNRAS,
300, 977
\bibitem[Hut 1997] {h97} Hut, P. Complexity, proceedings of the work shop on
fundemental sources of unpredictability, held at the Santa fr
institute, 3/96 
\bibitem[Ianna \etal 1987]{i87} Ianna P.A., Adler D.S., Faudree
E.F., 1987, AJ, 93, 347
\bibitem[Jeans 1929]{j29} Jeans J.H. Astronomy \& Cosmogony, 2nd ed., CUP.
\bibitem[Jones \& Shauffer 1991]{js91} Jones B.F., Shauffer J.R.,
1991, AJ, 102, 1080
\bibitem[King 1962]{k62} King I.R. 1962, A.J., 67, 471
\bibitem[Kroupa 2000]{k2000} Kroupa P. 2000 New Astronomy 4 615
\bibitem[Kroupa, Petr \& McCaughrean 1999]{kpmm99} Kroupa P., Petr
M.G., McCaughrean M.J., 1999 New Astronomy 4 1999
\bibitem[Kroupa 1995]{krou95} Kroupa P. 1995 MNRAS 277 1491
\bibitem[Kroupa, Tout \& Gilmore 1993]{ktg93} Kroupa P., Tout C.A. \&
Gilmore G. 1993 MNRAS 262 545
\bibitem[Leinert \etal 1993]{l93} Leinert Ch., Zinnecker H., Weitzel
N., Christou J., Ridgeway S.T., Jameson R., Haas M. \& Lenzen R. 1993,
A \& A, 278, 129 
\bibitem[Luhman \etal 2000]{luh2000} Luhman K.L., Reike G.H., Young
E.T., Cotera A.S., Chen H., Rieke M.J., Schneider G. \& Thompson
R.I. 2000 Ap. J. 540, 1016
\bibitem[Luhman \& Reike 1999]{lr99} Luhman K.L. \& Reike G.H. 1999
Ap.J. 497, 440
\bibitem[Luhman \& Reike 1998]{lr98} Luhman K.L. \& Reike G.H. 1998
Ap.J. 497, 354
\bibitem[Luhman \etal 1998]{letal98} Luhman K.R., Reike G.H., Lada
C.J. \& Lada E.A. 1998, Ap.J. 508, 347 
\bibitem[Mart\'{i}n \etal 2000]{m2000}  Mart\'{i}n E.L., Brandner
W., Bouvier J., Luhman K.L., Stauffer J., Basri G., Zapatero
Osorio M.R., Barrado Y Navascu\'{e}s D. 2000, ApJ, 543, 299
\bibitem[Mazeh \& Goldberg]{mg92} Mazeh T., Goldberg D. 1992, ApJ,
394, 592
\bibitem[Nakajima, T., Oppenheimer, B.R.,Kulkarni, S.R., Golimowski,
D.A.,Matthews, K. \& Durrance, S.T. 1995]{nak95} Nakajima T.,
Oppenheimer B.R., Kulkarni S.R., Golimowski D.A., Matthews K. \&
Durrance S.T. 1995, Nature, 378, 463
\bibitem[Perryman \etal 1998]{p98} Perryman M.A.C., Brown A.G.A.,
Lebreton Y., Gomez A., Turon C., De Strobel G.C., Mermilloid J.C.,
Robichon N., Kovalevsky J., Crifo F. 1998, A\&A, 331, 81
\bibitem[Pinfield \etal 2000]{p2000} Pinfield D.J., Hodgkin S.T.,
Jameson R.F., Cossburn M.R., Hambly N.C., 2000, MNRAS, 313, 347
\bibitem[Pinfield \etal 1998]{p1998} Pinfield D.J., Jameson R.F.,
Hodgkin S.T. 1998, MNRAS, 299, 955
\bibitem[Portegies Zwart \etal]{pz200} Portegies Zwart S.F.,
McMillan S.L.W., Hut P., Makino J., MNRAS in press
\bibitem[Rabound \& Mermilliod 1998] {rm98} Rabound D. \& Mermilliod
J.-C. 1998 A \& A 329 101
\bibitem[Reid \& Mahoney 2000]{rm2000} Reid I.N., Mahoney S., 2000,
MNRAS, 316, 827
\bibitem[Reid \& Hawley 1999]{rh99} Reid I.N., Hawley S.L. 1999, AJ,
117, 343
\bibitem[Richichi \etal 1994] {RLJ94} Richichi A., Leinhert Ch., Jameson R.,
Zinnecker H. 1994, A\&A, 287, 145
\bibitem[Sandrelli \etal 1999]{s99} Sandrelli S., Bragaglia A.,
Tosi M., Marconi G. 1999, MNRAS, 309, 739
\bibitem[Skrutskie \etal 1995]{sk1995}Skrutskie M.F., Beichman C.,
Capps R., Carpenter J., Chester T., Cutri R., Elias J., Elston
R., Huchra J.,Leibert J., Lonsdale C., Monet D., Price S.,
Scheider S., Seitzer P., Steining  R., Strom S., Weinberg M., 1995,
American Astronomical Society Meeting 187, \# 75.07 
\bibitem[Spitzer 1940]{sp40} Spitzer L. 1940 MNRAS 100 396
\bibitem[Stauffer, Hamilton \& Probst 1994]{shp94} Stauffer J.R.,
Hamilton D. \& ProbstR.G. 1994, AJ, 108, 155
\bibitem[Steele \& Jameson 1995] {SJ95} Steele I.A., Jameson R.F., 1995, MNRAS,
272, 630
\bibitem[Terlevich 1987] {t87} Terlevich E. 1987, MNRAS, 224, 193
\end{thebibliography}
\end{document}